\documentstyle[12pt]{article}
\global\arraycolsep=2pt %reduces the separation in eqnarrays

\newcommand{\AmS}{{\protect\the\textfont2
 A\kern-.1667em\lower.5ex\hbox{M}\kern-.125emS}}

\begin{document}
%%%%%%%%%%%%%%%%% New Commands (Nucl.\ Phys.\ ) %%%%%%%%%%%%%%%%%%%%%%%%%
\newcommand{\nn}{\noindent}
\newcommand{\cs}{\mbox{$\clubsuit$}}
\newcommand{\nl}{\nonumber \\}
\newcommand{\hf}{\hfill}
\newcommand{\naive}{na$\ddot{\imath}$ve}
\newcommand {\oa} {\mbox{${\cal O}( \alpha)$}}
\newcommand {\ho} {\mbox{${\cal O}( \alpha^{2})$}}
\hyphenation{brems-strah-lung}
%\documentstyle[12pt,epsfig,leqno]{article}
%\voffset1.2truecm
%\setlength{\unitlength}{1mm}
%\textwidth 16.0 true cm
% Increase texheight for the preprint numbers
%\textheight 23.0 true cm
%\headheight 0 cm
%\topmargin - 1.0 true cm
%\oddsidemargin 0.00 true in
%\renewcommand{\thefootnote}{\fnsymbol{footnote}}
%\baselineskip 1.4 cm
%
%%%%%%%%%%%%%%%%%%%%%%% NEWCOMMANDS (DAVID) %%%%%%%%%%%%%%%%%%%%%%%%%%%%
\def\ss{\footnotesize}
\def\SS{\footnotesize}
\def\sss{\scriptscriptstyle}
\def\barp{{\raise.35ex\hbox{${\sss (}$}}---{\raise.35ex\hbox{${\sss )}$}}}
\def\bdbarp{\hbox{$B_d$\kern-1.4em\raise1.4ex\hbox{\barp}}}
\def\bsbarp{\hbox{$B_s$\kern-1.4em\raise1.4ex\hbox{\barp}}}
\def\dbarp{\hbox{$D$\kern-1.1em\raise1.4ex\hbox{\barp}}}
\def\dcp{D^0_{\sss CP}}
\def\dbar{{\overline{D^0}}}
\newcommand{\xd}{x_d}
\newcommand{\xs}{x_s}
\newcommand{\bd}{B_d^0}
\newcommand{\bdb}{\overline{B_d^0}}
\newcommand{\bs}{B_s^0}
\newcommand{\bsb}{\overline{B_s^0}}
\newcommand{\bu}{B_u^\pm}
\newcommand{\beq}{\begin{equation}}
\newcommand{\eeq}{\end{equation}}
\newcommand{\absvcb}{\vert V_{cb}\vert}
\newcommand{\absvub}{\vert V_{ub}\vert}
\newcommand{\absvtd}{\vert V_{td}\vert}
\newcommand{\absvts}{\vert V_{ts}\vert}
\newcommand{\abseps}{\vert\epsilon\vert}
\newcommand{\epsp}{\epsilon^\prime/\epsilon}
\newcommand{\fbb}{f^2_{B_d}\hat{B}_{B_d}}
\newcommand{\fbbs}{f^2_{B_s}\hat{B}_{B_s}}
\newcommand{\fbd}{f_{B_d}}
\newcommand{\fbs}{f_{B_s}}
\newcommand{\fds}{f_{D_s}}
\def\rly#1{\mathrel{\raise.3ex\hbox{$#1$\kern-.75em\lower1ex\hbox{$\sim$}}}}
\def\lsim{\rly<}

% Journal and other miscellaneous abbreviations for references
\def \zpc#1#2#3{{\it Z.~Phys.,} C#1 (19#2) #3}
\def \plb#1#2#3{{\it Phys.~Lett.,} B#1 (19#2) #3}
\def \ibj#1#2#3{~#1, (19#2) #3}
\def \prl#1#2#3{{\it Phys.~Rev.~Lett.,} #1 (19#2) #3}
\def \prd#1#2#3{{\it Phys.~Rev.,} D#1 (19#2) #3} 
\def \npb#1#2#3{{\it Nucl.~Phys.}, B#1 (19#2) #3} 

\newread\epsffilein % file to \read
\newif\ifepsffileok % continue looking for the bounding box?
\newif\ifepsfbbfound % success?
\newif\ifepsfverbose % report what you're making?
\newdimen\epsfxsize % horizontal size after scaling
\newdimen\epsfysize % vertical size after scaling
\newdimen\epsftsize % horizontal size before scaling
\newdimen\epsfrsize % vertical size before scaling
\newdimen\epsftmp % register for arithmetic manipulation
\newdimen\pspoints % conversion factor
\pspoints=1bp % Adobe points are `big'
\epsfxsize=0pt % Default value, means `use natural size'
\epsfysize=0pt % ditto
\def\epsfbox#1{\global\def\epsfllx{72}\global\def\epsflly{72}%
 \global\def\epsfurx{540}\global\def\epsfury{720}%
 \def\lbracket{[}\def\testit{#1}\ifx\testit\lbracket
 \let\next=\epsfgetlitbb\else\let\next=\epsfnormal\fi\next{#1}}%
\def\epsfgetlitbb#1#2 #3 #4 #5]#6{\epsfgrab #2 #3 #4 #5 .\\%
 \epsfsetgraph{#6}}%
\def\epsfnormal#1{\epsfgetbb{#1}\epsfsetgraph{#1}}%
\def\epsfgetbb#1{%
%
% The first thing we need to do is to open the
% PostScript file, if possible.
%
\openin\epsffilein=#1
\ifeof\epsffilein\errmessage{I couldn't open #1, will ignore it}\else
%
% Okay, we got it. Now we'll scan lines until we find one that doesn't
% start with %. We're looking for the bounding box comment.
%
 {\epsffileoktrue \chardef\other=12
 \def\do##1{\catcode`##1=\other}\dospecials \catcode`\ =10
 \loop
 \read\epsffilein to \epsffileline
 \ifeof\epsffilein\epsffileokfalse\else
%
% We check to see if the first character is a % sign;
% if not, we stop reading (unless the line was entirely blank);
% if so, we look further and stop only if the line begins with
% `%%BoundingBox:'.
%
 \expandafter\epsfaux\epsffileline:. \\%
 \fi
 \ifepsffileok\repeat
 \ifepsfbbfound\else
 \ifepsfverbose\message{No bounding box comment in #1; using defaults}\fi\fi
 }\closein\epsffilein\fi}%
%
% Now we have to calculate the scale and offset values to use.
% First we compute the natural sizes.
%
\def\epsfclipstring{}% do we clip or not? If so,
\def\epsfclipon{\def\epsfclipstring{ clip}}%
\def\epsfclipoff{\def\epsfclipstring{}}%
\def\epsfsetgraph#1{%
 \epsfrsize=\epsfury\pspoints
 \advance\epsfrsize by-\epsflly\pspoints
 \epsftsize=\epsfurx\pspoints
 \advance\epsftsize by-\epsfllx\pspoints
%
% If `epsfxsize' is 0, we default to the natural size of the picture.
% Otherwise we scale the graph to be \epsfxsize wide.
%
 \epsfxsize\epsfsize\epsftsize\epsfrsize
 \ifnum\epsfxsize=0 \ifnum\epsfysize=0
 \epsfxsize=\epsftsize \epsfysize=\epsfrsize
 \epsfrsize=0pt
%
% We have a sticky problem here: TeX doesn't do floating point arithmetic!
% Our goal is to compute y = rx/t. The following loop does this reasonably
% fast, with an error of at most about 16 sp (about 1/4000 pt).
%
 \else\epsftmp=\epsftsize \divide\epsftmp\epsfrsize
 \epsfxsize=\epsfysize \multiply\epsfxsize\epsftmp
 \multiply\epsftmp\epsfrsize \advance\epsftsize-\epsftmp
 \epsftmp=\epsfysize
 \loop \advance\epsftsize\epsftsize \divide\epsftmp 2
 \ifnum\epsftmp>0
 \ifnum\epsftsize<\epsfrsize\else
 \advance\epsftsize-\epsfrsize \advance\epsfxsize\epsftmp \fi
 \repeat
 \epsfrsize=0pt
 \fi
 \else \ifnum\epsfysize=0
 \epsftmp=\epsfrsize \divide\epsftmp\epsftsize
 \epsfysize=\epsfxsize \multiply\epsfysize\epsftmp
 \multiply\epsftmp\epsftsize \advance\epsfrsize-\epsftmp
 \epsftmp=\epsfxsize
 \loop \advance\epsfrsize\epsfrsize \divide\epsftmp 2
 \ifnum\epsftmp>0
 \ifnum\epsfrsize<\epsftsize\else
 \advance\epsfrsize-\epsftsize \advance\epsfysize\epsftmp \fi
 \repeat
 \epsfrsize=0pt
 \else
 \epsfrsize=\epsfysize
 \fi
 \fi
%
% Finally, we make the vbox and stick in a \special that dvips can parse.
%
 \ifepsfverbose\message{#1: width=\the\epsfxsize, height=\the\epsfysize}\fi
 \epsftmp=10\epsfxsize \divide\epsftmp\pspoints
 \vbox to\epsfysize{\vfil\hbox to\epsfxsize{%
 \ifnum\epsfrsize=0\relax
 \includegraphics{#1}%
 \else
 \epsfrsize=10\epsfysize \divide\epsfrsize\pspoints
 \includegraphics{#1}%
 \fi
 \hfil}}%
\global\epsfxsize=0pt\global\epsfysize=0pt}%
%
% We still need to define the tricky \epsfaux macro. This requires
% a couple of magic constants for comparison purposes.
%
 {\catcode`\%=12 \global\let\epsfpercent=%\global\def\epsfbblit{%BoundingBox}}%
%
% So we're ready to check for `%BoundingBox:' and to grab the
% values if they are found.
%
\long\def\epsfaux#1#2:#3\\{\ifx#1\epsfpercent
 \def\testit{#2}\ifx\testit\epsfbblit
 \epsfgrab #3 . . . \\%
 \epsffileokfalse
 \global\epsfbbfoundtrue
 \fi\else\ifx#1\par\else\epsffileokfalse\fi\fi}%
%
% Here we grab the values and stuff them in the appropriate definitions.
%
\def\epsfempty{}%
\def\epsfgrab #1 #2 #3 #4 #5\\{%
\global\def\epsfllx{#1}\ifx\epsfllx\epsfempty
 \epsfgrab #2 #3 #4 #5 .\\\else
 \global\def\epsflly{#2}%
 \global\def\epsfurx{#3}\global\def\epsfury{#4}\fi}%
%
% We default the epsfsize macro.
%
\def\epsfsize#1#2{\epsfxsize}
%
% Finally, another definition for compatibility with older macros.
%
\let\epsffile=\epsfbox
%=======================================================================

%%%%%%%%%%%%%%%%%%%%%%% NEWCOMMANDS (AHMED) %%%%%%%%%%%%%%%%%%%%%%%%%%%%
\def\att{t \bar{t}}
\def\app{p \bar{p}}
\def\rts{\sqrt{s}}
\def\mt{m_t}
\def\mb{m_b}
\def\mc{m_c}
\newcommand{\bksgam}{\ $B \to K^*+ \gamma$}
\newcommand{\brogam}{\ $B \to \rho+ \gamma$}
\def\BDSl{B \to D^* \ell \nu_\ell}
\def\vdvp{v \cdot v^\prime}
\def\xiaoo{\xi_{A_1}(\vdvp =1 )}
\def\Vbc{V_{cb}}
\newcommand{\Tosc}{T_{osc}}
\newcommand{\sqrts}{\sqrt{s}}
\newcommand{\bg}{\beta \gamma}
\newcommand{\xds}{x_i}
\newcommand{\Ds}{D_s^\pm}
\newcommand{\bb}{B^0 B^0}
\newcommand{\barbar}{{\overline{B^0}}\thinspace{\overline{B^0}}}
\newcommand{\barb}{B^0 {\overline{B^0}}}
\newcommand{\bbar}{$B^0$-${\overline{B^0}}$}
\newcommand{\Deltat}{\Delta t}
\newcommand{\delt}{\delta t}
\newcommand{\delmd}{\Delta M_d}
\newcommand{\delms}{\Delta M_s}
\newcommand{\ps}{10^{-12} s}
\newcommand{\zbbar}{Z^0 \to b {\overline{b}}}
\newcommand{\eebbx}{$e^+ e^- \to B {\overline{B}} X$}
\newcommand{\pbpbbx}{$p{\overline{p}} \to B {\overline{B}} X$}
\newcommand{\kkbar}{$K^0$-${\overline{K^0}}$}
\newcommand{\bdbdbar}{$B_d^0$-${\overline{B_d^0}}$}
\newcommand{\bsbsbar}{$B_s^0$-${\overline{B_s^0}}$}
\newcommand{\as}{\mbox{$\alpha_{\displaystyle s}$}}
\newcommand{\aso}{\mbox{$O(\alpha_{\displaystyle s})$}}
\newcommand{\ass}{\mbox{$O(\alpha_{\displaystyle s}^2)$}}
\newcommand{\asq}{\mbox{$\alpha_{\displaystyle s}(Q^2)$}}
\newcommand{\ee}{\mbox{$e^+e^-$}}
\newcommand{\cc}{\mbox{$c {\overline{c}}$}}
\newcommand{\qq}{\mbox{$q {\overline{q}}$}}
\newcommand{\jp}{\mbox{$J/\Psi$}}
\newcommand{\lqc}{\Lambda_{QCD}}
\newcommand{\pmi}{{\not{p}}_{\perp}}
\newcommand{\set}{\sum E_{\perp}}
\newcommand{\ptr}{p_{\perp}}
\newcommand{\sww}{\sin^2{\theta_W}}
\newcommand{\sw}{\sin{\theta_W}}
%%%%%%%%%%%%%%%%%%%%%%%%%%%%%%%%%%%%%%%%%%%%%%%%%%%%%%%%%%%%%%%%%
\begin{flushright}
DESY 96-140 \\
UdeM-GPP-TH-96-38\\
%hep-ph/9907yyy \\
July 1996\\
\end{flushright}
\begin{center}
{\Large \bf
\centerline{CP Violation and Flavour Mixing}
\centerline{in the Standard Model -- 1996 Update}}
\vspace*{1.5cm}
%\vskip1cm
 {\large A.~Ali}$\footnote{Presented at the QCD Euroconference 96,
Montpellier, July 4 - 12, 1996.}$
\vskip0.2cm
 Deutsches Elektronen Synchrotron DESY, Hamburg \\
%\vspace*{0.5cm}
\vspace*{0.3cm}
\centerline{ and}
%\vspace*{0.5cm}
\vspace*{0.3cm}
{\large D.~London}$\footnote{Permanent address: Laboratoire de
physique nucl\'eaire, Universit\'e de Montr\'eal, C.P. 6128, succ.\
centre-ville, Montr\'eal, QC, Canada H3C 3J7.}$ \\
\smallskip
%\vspace*{0.5cm}
Physics Department, McGill University \\
3600 University St., Montr\'eal, Qu\'ebec, Canada, H3A 2T8\\
%\vskip1cm
\vskip0.5cm
{\Large Abstract\\}
\parbox[t]{\textwidth}{
\indent
We review and update the constraints on the parameters of the quark flavour
mixing matrix $V_{CKM}$ in the standard model and estimate the resulting CP
asymmetries in $B$ decays, taking into account recent experimental and
theoretical developments. With the
updated CKM matrix we present the currently-allowed range of the ratios
$|V_{td}/V_{ts}|$ and $|V_{td}/V_{ub}|$, as well as the standard model
predictions for the \bsbsbar\ mixing parameter $\xs$ (or, equivalently,
$\delms$) and the quantities $\sin 2\alpha$, $\sin 2\beta$ and
$\sin^2\gamma$, which characterize the CP-asymmetries in $B$-decays.
}
%\vskip2cm
\end{center}
\thispagestyle{empty}
\newpage
\setcounter{page}{1}
% Decrease texheight (for preprint numbers) again
\textheight 23.0 true cm

%%%%%%%%%%%%%%%%%%%%%% SECTION 1 %%%%%%%%%%%%%%%%%%%%%%%%%%%%%%%%

\section{An Update of the CKM Matrix}

We revise and update the profile of the Cabibbo-Kobayashi-Maskawa (CKM)
matrix \cite{CKM} reported by us in 1995 \cite{AL95}. In particular, we
focus on the CKM unitarity triangle and CP asymmetries in $B$ decays, which
are the principal objects of interest in experiments at present and 
forthcoming $B$ facilities. In performing this update, we include the
improvements reported in a number of measurements of the lifetime, mixing
ratio, and the CKM matrix elements $\absvcb$ and $\vert V_{ub}/V_{cb}
\vert$ from $B$ decays, as well as the top quark mass, $\abseps$, and
progress in theoretical calculations involving a number of perturbative and
non-perturbative aspects of QCD.

In updating the CKM matrix elements, we make use of the Wolfenstein
parametrization \cite{Wolfenstein}, which follows from the observation that
the elements of this matrix exhibit a hierarchy in terms of $\lambda$, the
Cabibbo angle. In this parametrization the CKM matrix can be written
approximately as
\beq
V_{CKM} \simeq \left(\matrix{
 1-{1\over 2}\lambda^2 & \lambda
 & A\lambda^3 \left( \rho - i\eta \right) \cr
 -\lambda ( 1 + i A^2 \lambda^4 \eta )
& 1-{1\over 2}\lambda^2 & A\lambda^2 \cr
 A\lambda^3\left(1 - \rho - i \eta\right) & -A\lambda^2 & 1 \cr}\right)~.
\label{CKM}
\eeq
We shall discuss those quantities which constrain these CKM parameters,
pointing out the significant changes in the determination of $\lambda$,
$A$, $\rho$ and $\eta$, as compared to \cite{AL95}. Also, for reasons of
brevity, we shall be rather concise in this report and refer to
\cite{AL94,AL95} for further details. 

We recall that $\vert V_{us}\vert$ has been extracted with good accuracy 
from $K\to\pi e\nu$ and hyperon decays \cite{PDG96} to be
\beq
\vert V_{us}\vert=\lambda=0.2205\pm 0.0018~.
\eeq
This agrees quite well with the determination of $V_{ud}\simeq 1-{1\over
2}\lambda^2$ from $\beta$-decay \cite{PDG96},
\beq
\vert V_{ud}\vert=0.9736\pm 0.0010~.
\eeq

We note and comment on the changes that we have made in the input to our
present analysis compared to that reported by us last year in
Ref.~\cite{AL95}:
\begin{itemize}

\item {$\boldmath m_t$}: The present average of the top quark mass measured 
directly at Fermilab by the CDF and D0 collaborations is $\mt =175 \pm 9
$ GeV \cite{DPG96}. We interpret this as being the pole mass (though this
identification is not unambiguous). This, in turn, leads to the running
top quark mass in the $\overline{MS}$ scheme, $\overline{\mt} (\mt)= 165
\pm 9 $ GeV \cite{mtmsbar}.

\item {$\boldmath \absvcb $}: The determination of $\absvcb$ from 
inclusive and exclusive $B$ decays has been reviewed earlier in a number of
studies \cite{AL95,neubert95,shifman95,Tomasz95}. Here, we shall 
concentrate on the exclusive decay $B \to D^* \ell \nu_\ell$ analyzed in
the context of heavy quark effective theory (HQET), as this method seems to
have been scrutinized in great detail. Using HQET, the differential decay
rate in $B \to D^* \ell \nu_\ell$ is \cite{Wisgur}
\begin{eqnarray}
\frac{d\Gamma (B \to D^* \ell \bar{\nu})}{d\omega }
&=& \frac{G_F^2}{48 \pi^3} (m_B-m_{D^*})^2 m_{D^*}^3 \eta_{A}^2
 \sqrt{\omega^2-1} (\omega + 1)^2 \\ \nonumber
&~& ~~~~~~~~\times [ 1+ \frac{4 \omega}{\omega + 1}
 \frac{1-2\omega r + r^2}{(1-r)^2}] \absvcb ^2 \xi^2(\omega) ~,
\label{bdstara1}
\end{eqnarray}
where $r=m_{D^*}/m_B$, $\omega=v\cdot v'$ ($v$ and $v'$ are the
four-velocities of the $B$ and $D^*$ meson, respectively), and $\eta_{A}$
is the short-distance correction to the axial vector form factor.
Measurements of the intercept ${\cal F}(1) \vert V_{cb} \vert$ (with ${\cal
F}(\omega) \equiv \eta_A \cdot \xi(\omega)$) in the decays $B \to D^* \ell
\nu_\ell$, have been reported by the ALEPH, ARGUS, DELPHI, CLEO, and OPAL
collaborations, and a careful job of averaging the experimental results on
the intercept ${\cal F}(1)\absvcb$ has been performed for the presently
available data by Gibbons \cite{Gibbons96}. The intercept is highly
correlated with the slope of the Isgur-Wise function ${\cal F}(y)$
\cite{Wisgur} measured in each experiment and a simultaneous average of
the slope and the intercept is required with the correlations included. We
refer to Ref.~\cite{Gibbons96} for the details of the analysis, and quote
the final result, which reads as
\begin{equation}
 {\cal F}(1) \absvcb = 0.0357 \pm 0.0020 \pm 0.0014,
\label{gibbonsf1vcb}
\end{equation}
where the first error is statistical plus systematic, and the second is the 
estimate of the curvature bias in extrapolating the function ${\cal F}(y)$.
Theoretical estimates for the quantity ${\cal F}(1)$ are on a firmer
footing, as the QCD perturbative part $\eta_A$ has been calculated at
next-to-leading order in Ref.~\cite{Cz96}, yielding $\eta_A=0.965 \pm
0.007 + {\cal O}(\as^3)$, which reduces the perturbative QCD error on
$\eta_A$ by a factor 3, as compared to the earlier estimates of the same.
(See, for example, the work by M. Neubert \cite{neubert95}.) The remaining
theoretical uncertainty is now in the power corrections to the Isgur-Wise
function at the symmetry point $\xi(1)$. The leading power corrections to
$\xi(1)$ are absent due to Luke's theorem \cite{Luke}. There exist
extensive studies of the $1/M_Q^2$ corrections and we refer to
\cite{neubert95,shifman95,GKLW96} for detailed discussions of the
corrections and their model dependence. Using these estimates gives: 
\begin{equation}
\label{czarneskixi}
\xi (1) = 1+ \delta (1/m^2)= 0.945 \pm 0.025 ~
\Longrightarrow {\cal F}(1)=0.907 \pm 0.026~.
\end{equation}
The present theoretical error $\Delta {\cal F}(1)/{\cal F}(1)= 0.029$
(which appears to us to be rather irreducible) is about $2/3$ the size of
the theoretical error used in our earlier CKM fits \cite{AL95}, where we
had used ${\cal F}(1)=0.91 \pm 0.04$. Taking into account the updated
experimental \cite{Gibbons96} and theoretical \cite{Cz96} input, the
present determination of $\absvcb$ is:
\beq
\vert V_{cb} \vert= 0.0393 \pm 0.0021 ~(expt) \pm 0.0015 ~(curv)
\pm 0.0011~(th),
\label{Vcbhqet95}
\eeq
In the fits below we have added the errors in quadrature, getting
\beq 
 \vert V_{cb} \vert = 0.0393 \pm 0.0028~,
\label{vcbnow}
\eeq
yielding
\beq
A = 0.81 \pm 0.058~.
\label{Avalue}
\eeq
This represents a measurement of this parameter at $\pm 7.2\%$, making it
after $\lambda$ the next best determined CKM parameter. We note that the 
central value of $A$ is essentially the same as that used in \cite{AL95},
but the error is now reduced. 

\item {$ \boldmath |V_{ub}/V_{cb}| $}: The knowledge of the CKM matrix
element ratio $|V_{ub}/V_{cb}|$ is based on the analysis of the end-point
lepton energy spectrum in semileptonic decays $B \to X_{u} \ell \nu_\ell$
and the measurement of the exclusive semileptonic decays $B \to (\pi, \rho)
\ell \nu_\ell$ reported by the CLEO collaboration \cite{Tomasz95}. As noted
in \cite{Bartelt93}, the inclusive measurements suffer from a large 
extrapolation factor from the measured end-point rate to the total
branching ratio, which is model dependent. The exclusive measurements allow
a discrimination among a number of models \cite{Tomasz95}, all of which
were previously allowed from the inclusive decay analysis alone. It is
difficult to combine the exclusive and inclusive measurements to get a
combined determination of $\absvub/\absvcb$. However, it has been noted
that the disfavoured models in the context of the exclusive decays are also
those which introduce a larger theoretical dispersion in the interpretation
of the inclusive $B \to X_u \ell \nu_\ell$ data. Excluding them from
further consideration, measurements in both the inclusive and exclusive
modes are compatible with \cite{Gibbons96}:
\beq
\left\vert \frac{V_{ub}}{V_{cb}} \right\vert = 0.08\pm 20\%~.
\label{vubvcbn}
\eeq
This gives
\beq
\sqrt{\rho^2 + \eta^2} = 0.363 \pm 0.073~.
\eeq
Again, the central value of this quantity is the same as that used by us
in \cite{AL95}, but the error is marginally reduced.

\item {$\boldmath \abseps, \hat{B}_K, ~\mbox{and constraints on} ~\rho 
~\mbox{and} ~\eta$}: The experimental value of $\abseps$ has changed
somewhat from our previous analyses, and the error has decreased
\cite{PDG96}:
\beq
\abseps = (2.280\pm 0.013)\times 10^{-3}~.
\eeq
Theoretically, $\abseps$ is essentially proportional to the imaginary part
of the box diagram for \kkbar\ mixing and is given by \cite{Burasetal}
\begin{eqnarray}
\abseps &=& \frac{G_F^2f_K^2M_KM_W^2}{6\sqrt{2}\pi^2\Delta M_K}
\hat{B}_K\left(A^2\lambda^6\eta\right)
\bigl(y_c\left\{\hat{\eta}_{ct}f_3(y_c,y_t)-\hat{\eta}_{cc}\right\}
 \nonumber \\
&~& ~~~~~~~~~~~~~~+ 
~\hat{\eta}_{tt}y_tf_2(y_t)A^2\lambda^4(1-\rho)\bigr), 
\label{eps}
\end{eqnarray}
where $y_i\equiv m_i^2/M_W^2$, and the functions $f_2$ and $f_3$ can be
found in Ref.~\cite{AL94}. Here, the $\hat{\eta}_i$ are QCD correction
factors, calculated at next-to-leading order in \cite{HN94}
($\hat{\eta}_{cc}$), \cite{etaB} ($\hat{\eta}_{tt}$) and \cite{HN95}
($\hat{\eta}_{ct}$). The theoretical uncertainty in the expression for 
$\abseps$ is in the renormalization-scale independent parameter
$\hat{B}_K$, which represents our ignorance of the hadronic matrix
element $\langle K^0 \vert {({\overline{d}}\gamma^\mu (1-\gamma_5)s)}^2
\vert {\overline{K^0}}\rangle$. Some recent calculations of $\hat{B}_K$
using lattice QCD methods \cite{Soni95} and the $1/N_c$ approach
\cite{BP95} are: $\hat{B}_K=0.83 \pm 0.03$ (Sharpe \cite{Sharpe94}),
$\hat{B}_K=0.86 \pm 0.15$ (APE Collaboration \cite{Crisafulli95}),
$\hat{B}_K=0.67 \pm 0.07$ (JLQCD Collaboration \cite{JLQCD}), 
$\hat{B}_K=0.78 \pm 0.11$ (Bernard and Soni \cite{JLQCD}), and
$\hat{B}_K=0.70 \pm 0.10$ (Bijnens and Prades \cite{BP95}). They strongly
suggest that the theoretical dispersion on this quantity has been greatly
reduced compared to the ranges $\hat{B}_K=0.8 \pm 0.2$ and $\hat{B}_K=0.6
\pm 0.2$, which we had used previously as the best estimates in the
Lattice-QCD and chiral perturbation theory frameworks, respectively. The
more recent calculations given above are compatible with the range
\begin{equation}
 \hat{B}_K=0.75 \pm 0.10 ,
\label{BKrange}
\end{equation}
which we now use in our analysis. This is one of the principal sources of
reduction in the allowed values of $\rho$ and $\eta$, as we shall see
later.

\item {$\boldmath \Delta M_d, \fbb, ~\mbox{and constraints on} ~\rho
~\mbox{and} ~\eta$}: From a theoretical point of view we prefer to use the
mass difference $\Delta M_d$, as it liberates one from the errors on the
lifetime measurement. The present world average for $\Delta M_d$ is
\cite{Gibbons96}
\beq
%\Delta M_d = 0.47 \pm 0.017~(ps)^{-1} ~.
\Delta M_d = 0.464 \pm 0.018~(ps)^{-1} ~.
\label{deltamd}
\eeq
The mass difference $\Delta M_d$ is calculated from the \bdbdbar\ box
diagram. Unlike the kaon system, where the contributions of both the $c$-
and the $t$-quarks in the loop are important, this diagram is dominated by
$t$-quark exchange:
\beq
\label{bdmixing}
\Delta M_d = \frac{G_F^2}{6\pi^2}M_W^2M_B\left(\fbb\right)\hat{\eta}_B y_t
f_2(y_t) \vert V_{td}^*V_{tb}\vert^2~, \label{xd}
\eeq
where, using Eq.~\ref{CKM}, $\vert V_{td}^*V_{tb}\vert^2= A^2\lambda^{6}
[\left(1-\rho\right)^2+\eta^2]$. Here, $\hat{\eta}_B$ is the QCD
correction. In Ref.~\cite{etaB}, this correction was analyzed including the
effects of a heavy $t$-quark. It was found that $\hat{\eta}_B$ depends
sensitively on the definition of the $t$-quark mass, and that, strictly
speaking, only the product $\hat{\eta}_B(y_t)f_2(y_t)$ is free of this
dependence. In the fits presented here we use the value
$\hat{\eta}_B=0.55$, calculated in the $\overline{MS}$ scheme, following
Ref.~\cite{etaB}. Consistency requires that the top quark mass be rescaled
from its pole (mass) value of $\mt =175 \pm 9$ GeV to the value
$\overline{\mt}(\mt(pole))$ in the $\overline{MS}$ scheme, given above.

For the $B$ system, the hadronic uncertainty is given by $\fbb$, analogous
to $\hat{B}_K$ in the kaon system, except that in this case $\fbd$ has not
been measured. The present status of the lattice-QCD estimates for $\fbd$,
$\hat{B}_{B_d}$ and related quantities for the $B_s$ meson, obtained in the
quenched (now usually termed as the valence) approximation was summarized
recently in \cite{Wittig96}, giving
\begin{eqnarray}
 \fbd &=& 170 ^{+55}_{-50} ~\mbox{MeV}, \nonumber\\
 \hat{B}_{B_d} &=& 1.02 ^{+0.05 ~~+0.03}_{-0.06 ~~-0.02},
\end{eqnarray}
where the first error on $\hat{B}_{B_d}$ is statistical and the second 
systematic, estimated by the UKQCD collaboration \cite{UKQCDBB}. The effect
of unquenching is an estimated $10\%$ increase in the value of $\fbd$
(likewise $\fbs$) \cite{Petronzio96}. A modern estimate of $\fbb$ in the
QCD sum rule approach is that given in \cite{narison95}, which is
stated in terms of $f_\pi$, and on using $f_\pi =132$ MeV translates into
\begin{equation}
\fbd \sqrt{\hat{B}_{B_d}} = 197 \pm 18 ~\mbox{MeV}~,
\label{FBrangesr}
\end{equation} 
In our fits, we will take 
\begin{equation}
\fbd \sqrt{\hat{B}_{B_d}} = 200 \pm 40 ~\mbox{MeV}~,
\label{FBrange}
\end{equation}
which is compatible with the results from both lattice-QCD and QCD sum
rules for this quantity. We note that the range (\ref{FBrange}) is
considerably tighter as compared to our previous theoretical estimates of
the same, namely $\fbd \sqrt{\hat{B}_{B_d}} = 180 \pm 53 ~\mbox{MeV}$. This
is the second most important source of reduction in the allowed CKM
parameter space.

\item {$ \boldmath \delms, ~\delms /\delmd ~\mbox{and constraints on the
unitarity triangle}$}: We also estimate the SM prediction for the \bsbsbar\
mixing parameter, $\delms$ and $\xs$. The experimental lower limits on
these quantities have steadily increased, thanks to the experiments at LEP
\cite{Saulanwu95,ALEPHxs96} and more recently also from SLC. These limits
have now started to be significant for the allowed CKM parameter space.
The present best limit from a single experiments comes from 
ALEPH,  $\delms > 7.8$ (ps)$^{-1}$ \cite{ALEPHxs96,Zeitnitz}. However,
combining this with the corresponding limit from  
DELPHI yields $\delms > 9.2$ (ps)$^{-1}$ \cite{Gibbons96},
 leading to
$\delms /\delmd > 19.0$ (95\% C.L.). We show how this limit constrains
the CKM parameter space. We also give the present $95\%$ C.L. upper and lower
bounds on the matrix element ratio $\vert V_{td}/V_{ts} \vert$, as well as
the allowed (correlated) values of the CKM matrix elements $|V_{td}|$ and
$|V_{ub}|$.
\end{itemize}

As in our previous analysis, we consider two types of fits. In Fit 1, we
assume particular fixed values for the theoretical hadronic quantities. The
allowed ranges for the CKM parameters are derived from the (Gaussian)
errors on experimental measurements only. In Fit 2, we assign a central
value plus an error (treated as Gaussian) to the theoretical quantities. In
the resulting fits, we combine the experimental and theoretical errors in
quadrature. For both fits we calculate the allowed region in CKM parameter
space at 95\% C.L.

We also present the corresponding allowed ranges for the CP-violating
phases that will be measured in $B$ decays, characterized by $\sin 2\beta$,
$\sin 2\alpha$ and $\sin^2\gamma$. These can be measured directly through
rate asymmetries in the decays $\bdbarp \to J/\psi K_S$, $\bdbarp \to \pi^+
\pi^-$, and $\bsbarp\ \to D_s^\pm K^\mp$ (or $B^\pm \to \dbarp\ K^\pm$),
respectively. We also show the allowed correlated domains for two of the CP
asymmetries $(\sin 2\alpha,\sin 2\beta)$, as well as the correlation
between $\alpha$ and $\gamma$. 

This paper is organized as follows. In Section 2, we present the results of
our updated fits for the CKM parameters. These results are summarized in
terms of the allowed domains of the unitarity triangle, which are displayed
in several figures and tables. In Section 3, we discuss the impact of the
recent lower limit on the ratio $\Delta M_s/\Delta M_d$ reported by the
ALEPH collaboration on the CKM parameters and estimate the expected range
of the mixing ratio $\xs$ and $\delms$ in the SM based on our fits. Here we 
also present the allowed 95\% C.L. range for $\vert V_{td}/V_{ts} \vert$.
In Section 4 we discuss the predictions for the CP asymmetries in the
neutral $B$ meson sector and calculate the correlations for the CP
violating asymmetries proportional to $\sin 2\alpha$, $\sin 2 \beta$ and
$\sin^2 \gamma$. We present here the allowed values of the CKM matrix
elements $|V_{td}|$ and $|V_{ub}|$. Section 5 contains a summary and an
outlook for improving the profile of the CKM unitarity triangle.

\begin{table}
\hfil
\vbox{\offinterlineskip
\halign{&\vrule#&
 \strut\quad#\hfil\quad\cr
\noalign{\hrule}
height2pt&\omit&&\omit&\cr
& Parameter && Value & \cr
height2pt&\omit&&\omit&\cr
\noalign{\hrule}
height2pt&\omit&&\omit&\cr
& $\lambda$ && $0.2205$ & \cr
& $\vert V_{cb} \vert $ && $0.0393 \pm 0.0028$ & \cr
& $\vert V_{ub} / V_{cb} \vert$ && $0.08 \pm 0.016$ & \cr
& $\abseps$ && $(2.280 \pm 0.013) \times 10^{-3}$ & \cr
& $\Delta M_d$ && $(0.464 \pm 0.018)~(ps)^{-1}$ & \cr 
& $\tau(B_d)$ && $(1.55 \pm 0.04)~(ps)$ & \cr
& $\overline{\mt}(\mt(pole))$ && $(165 \pm 9)$ GeV & \cr
& $\hat{\eta}_B$ && $0.55$ & \cr
& $\hat{\eta}_{cc} $ && $1.38$ & \cr
& $\hat{\eta}_{ct} $ && $0.47$ & \cr
& $\hat{\eta}_{tt} $ && $0.57$ & \cr
& $\hat{B}_K$ && $0.75 \pm 0.1$ & \cr
& $\fbd\sqrt{\hat{B}_{B_d}} $ && $200 \pm 40$ MeV & \cr
height2pt&\omit&&\omit&\cr
\noalign{\hrule}}}
\caption{Parameters used in the CKM fits. Values of the hadronic quantities
$\hat{B}_K$ and $\fbd \protect\sqrt{\hat{B}_{B_d}}$ shown are motivated by 
lattice QCD results, QCD sum rules and chiral perturbation theory. In Fit
1, specific values of these hadronic quantities are chosen, while in Fit 2,
they are allowed to vary over the given ranges.}
\label{tabfit}
\end{table}

%%%%%%%%%%%%%%%%%%%%%% SECTION 2 %%%%%%%%%%%%%%%%%%%%%%%%%%%%%%%%

\section{The Unitarity Triangle}

The allowed region in $\rho$-$\eta$ space can be displayed quite elegantly
using the so-called unitarity triangle. The unitarity of the CKM matrix
leads to the following relation:
\beq
V_{ud} V_{ub}^* + V_{cd} V_{cb}^* + V_{td} V_{tb}^* = 0~.
\eeq
Using the form of the CKM matrix in Eq.~\ref{CKM}, this can be recast as
\beq
\label{trianglerel}
\frac{V_{ub}^*}{\lambda V_{cb}} + \frac{V_{td}}{\lambda V_{cb}} = 1~,
\eeq
which is a triangle relation in the complex plane (i.e.\ $\rho$-$\eta$
space), illustrated in Fig.~\ref{triangle}. Thus, allowed values of $\rho$
and $\eta$ translate into allowed shapes of the unitarity triangle.

% This is Figure 1
\begin{figure}
%\centerline{\psfig{figure=rhoeta.ps,height=5.0cm,angle=90}}
%\epsfig{file=rhoeta1.ps,bbllx=30pt,bblly=285pt,bburx=390pt,bbury=494pt,
%width=10cm}
\vskip -1.0truein
\centerline{\epsfxsize 3.5 truein \epsfbox {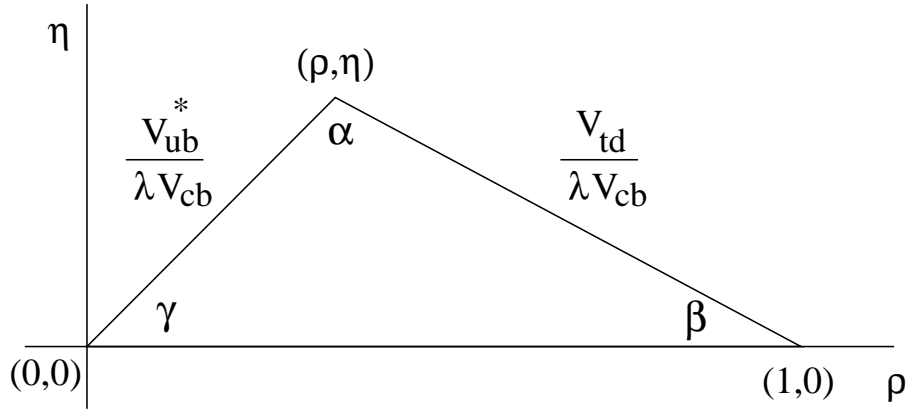}}
\vskip -1.2truein
\caption{The unitarity triangle. The angles $\alpha$, $\beta$ and $\gamma$
can be measured via CP violation in the $B$ system.}
\label{triangle}
\end{figure}

In order to find the allowed unitarity triangles, the computer program
MINUIT is used to fit the CKM parameters $A$, $\rho$ and $\eta$ to the
experimental values of $\absvcb$, $\vert V_{ub}/V_{cb}\vert$, $\abseps$ and
$\xd$. Since $\lambda$ is very well measured, we have fixed it to its
central value given above. As discussed in the introduction, we present
here two types of fits:
\begin{itemize}
\item
Fit 1: the ``experimental fit.'' Here, only the experimentally measured
numbers are used as inputs to the fit with Gaussian errors; the coupling
constants $f_{B_d} \sqrt{\hat{B}_{B_d}}$ and $\hat{B}_K$ are given fixed
values.
\item
Fit 2: the ``combined fit.'' Here, both the experimental and theoretical
numbers are used as inputs assuming Gaussian errors for the theoretical
quantities.
\end{itemize}

We first discuss the ``experimental fit" (Fit 1). The goal here is to
restrict the allowed range of the parameters ($\rho,\eta)$ for given values
of the coupling constants $f_{B_d} \sqrt{\hat{B}_{B_d}}$ and $\hat{B}_K$.
For each value of $\hat{B}_K$ and $f_{B_d}\sqrt{\hat{B}_{B_d}}$, the CKM
parameters $A$, $\rho$ and $\eta$ are fit to the experimental numbers given
in Table \ref{tabfit} and the $\chi^2$ is calculated.
\begin{table}
\hfil
\vbox{\offinterlineskip
\halign{&\vrule#&
 \strut\quad#\hfil\quad\cr
\noalign{\hrule}
height2pt&\omit&&\omit&&\omit&\cr
& $\fbd\sqrt{\hat{B}_{B_d}}$ (MeV) && $(\rho,\eta)$ && $\chi^2_{min}$ & \cr
height2pt&\omit&&\omit&&\omit&\cr
\noalign{\hrule}
height2pt&\omit&&\omit&&\omit&\cr
& $130$ && $(-0.37,~0.20)$ && $2.53$ & \cr
& $140$ && $(-0.32,~0.24)$ && $0.94$ & \cr
& $150$ && $(-0.27,~0.27)$ && $0.23$ & \cr
& $160$ && $(-0.21,~0.30)$ && $1.4 \times 10^{-2}$ & \cr
& $170$ && $(-0.15,~0.33)$ && $1.0 \times 10^{-3}$ & \cr
& $180$ && $(-0.08,~0.35)$ && $2.8 \times 10^{-2}$ & \cr
& $190$ && $(0.0,~0.36)$ && $7.9 \times 10^{-3}$ & \cr
& $200$ && $(0.07,~0.36)$ && $1.6 \times 10^{-2}$ & \cr
& $210$ && $(0.13,~0.37)$ && $0.20$ & \cr
& $220$ && $(0.19,~0.37)$ && $0.70$ & \cr
& $230$ && $(0.24,~0.37)$ && $1.61$ & \cr
& $240$ && $(0.29,~0.37)$ && $2.97$ & \cr
height2pt&\omit&&\omit&&\omit&\cr
\noalign{\hrule}}}
\caption{The ``best values'' of the CKM parameters $(\rho,\eta)$ as a
function of the coupling constant $\fbd\protect\sqrt{\hat{B}_{B_d}}$,
obtained by a minimum $\chi^2$ fit to the experimental data. We fix
$\hat{B}_K=0.75$. The resulting minimum $\chi^2$ values from the MINUIT
fits are also given.} 
\label{chifitab}
\end{table}

First, we fix $\hat{B}_K = 0.75$, and vary $f_{B_d}\sqrt{\hat{B}_{B_d}}$ in
the range 130 MeV to 240 MeV. The resulting $\chi^2_{min}$ values are given
in Table \ref{chifitab}, together with the best-fit values of the CKM
parameters $(\rho,\eta)$. We note that for this value of $\hat{B}_K$, 
certain values of $f_{B_d}\sqrt{\hat{B}_{B_d}}$ are disfavoured since they
do not provide a good fit to the data. Since we have two variables ($\rho$
and $\eta$), we use $\chi^2_{min}<2.0$ as our ``good fit" criterion, and
we find that $f_{B_d} \sqrt{\hat{B}_{B_d}}\leq 130$ MeV and $f_{B_d}
\sqrt{\hat{B}_{B_d}} \geq 240$ MeV give poor fits to the existing data. As
the allowed range of $f_{B_d}\sqrt{\hat{B}_{B_d}}$ from this fit has a
large overlap with the theoretically motivated range given above, we note
that present data do not allow a further reduction of this uncertainty. The
fits are presented as an allowed region in $\rho$-$\eta$ space at 95\% C.L.
($\chi^2 = \chi^2_{min} + 6.0$). The results are shown in
Fig.~\ref{rhoeta1}. As we pass from Fig.~\ref{rhoeta1} (top left) to
Fig.~\ref{rhoeta1} (bottom right), the unitarity triangles represented by 
these graphs
become more and more obtuse. However, the range of possibilities for these
triangles is now considerably reduced as compared to the earlier fits we
have presented in \cite{AL95}. This is due in part to the (somewhat)
improved measurements of $\absvcb$ and $|V_{ub}/V_{cb}|$, but mainly
reflects our reduced theoretical errors on the quantities $\hat{B}_K$ and
$\fbb$. We hope that this trust in the improved calculational ability of
these parameters is well placed! There are two things to be learned from
this fit. First, our quantitative knowledge of the unitarity triangle is
at present not very solid. This will be seen more clearly when we present
the results of Fit 2. Second, unless our knowledge of hadronic matrix
elements improves considerably, measurements of $\abseps$ and $x_d$, no
matter how precise, will not help much in further constraining the
unitarity triangle. This is why measurements of CP-violating rate
asymmetries in the $B$ system are so important \cite{BCPasym,AKL94}. Being
largely independent of theoretical uncertainties, they will allow us to
accurately pin down the unitarity triangle. With this knowledge, we could
deduce the correct values of $\hat{B}_K$ and $f_{B_d}\sqrt{\hat{B}_{B_d}}$,
and thus rule out or confirm different theoretical approaches to
calculating these hadronic quantities.

% This is Figure 2
\begin{figure}
\vskip -2.4truein
\centerline{\epsfxsize 7.0 truein \epsfbox {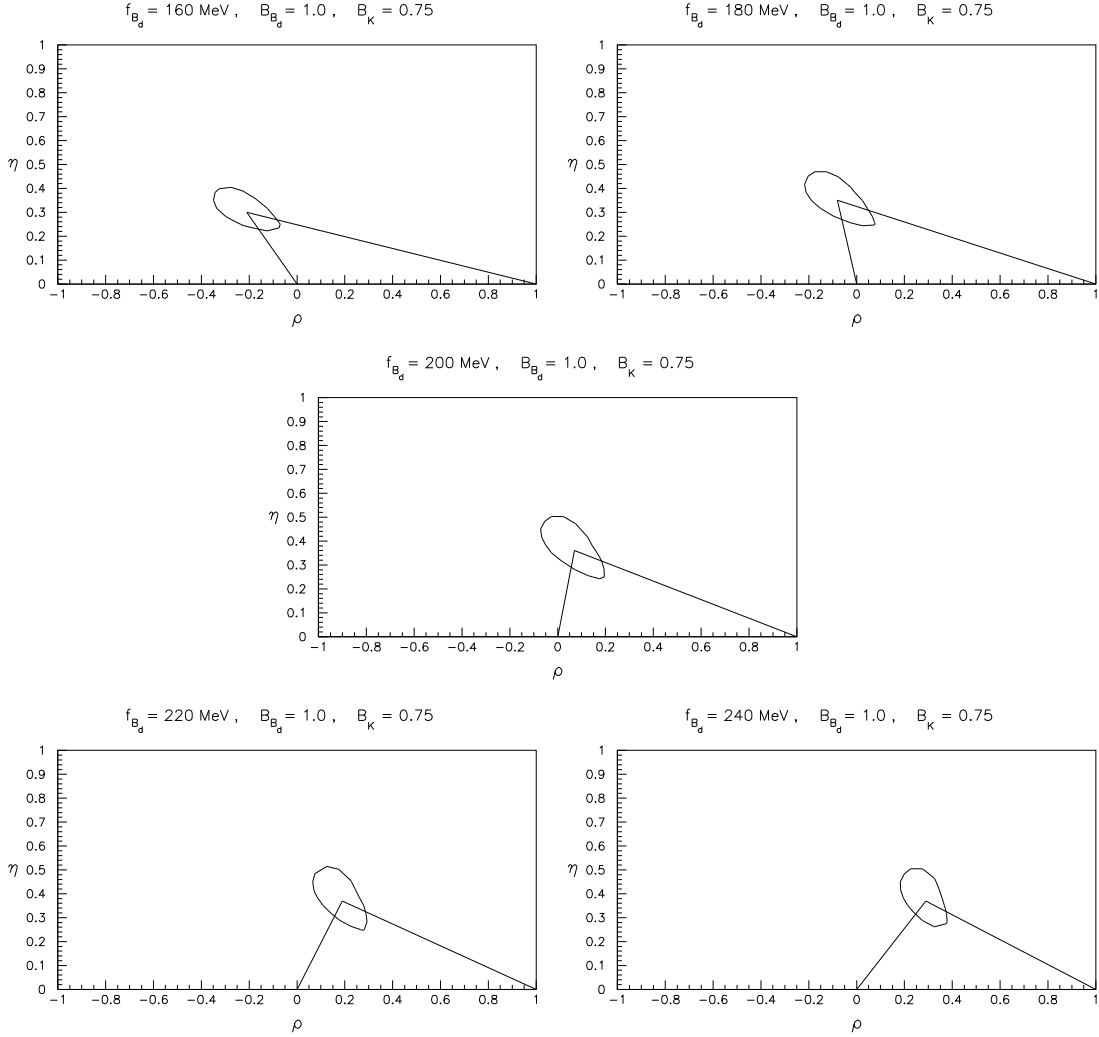}}
\vskip -1.0truein
\caption{Allowed region in $\rho$-$\eta$ space, from a fit to the
experimental values given in Table \protect{\ref{tabfit}}. We have fixed
$\hat{B}_K=0.75$ and vary the coupling constant product
$\fbd\protect\sqrt{\hat{B}_{B_d}}$ as indicated on the figures. The solid
line represents the region with $\chi^2=\chi_{min}^2+6$ corresponding to
the 95\% C.L.\ region. The triangles show the best fit.}
\label{rhoeta1}
\end{figure}

We now discuss the ``combined fit" (Fit 2). Since the coupling constants
are not known and the best we have are estimates given in the ranges in
Eqs.~(\ref{BKrange}) and (\ref{FBrange}), a reasonable profile of the
unitarity triangle at present can be obtained by letting the coupling
constants vary in these ranges. The resulting CKM triangle region is shown
in Fig.~\ref{rhoeta2}. As is clear from this figure, the allowed region is
still rather large at present. However, present data and theory do
restrict the parameters $\rho$ and $\eta$ to lie in the following
range:
\begin{eqnarray}
 0.20 &\leq & \eta \leq 0.52 , \nonumber \\
 -0.35 &\leq & \rho \leq 0.35 ~.
\label{rhoetarange}
\end{eqnarray}
The preferred values obtained from the ``combined fit" are
\beq
(\rho,\eta) = (0.05,0.36) ~~~(\mbox{with}~\chi^2 = 6.6\times 10^{-3})~,
\eeq
which gives rise to an almost right-angled unitarity triangle, with the
angle $\gamma$ being close to $90$ degrees. However, as we quantify below,
the allowed ranges of the CP violating angles $\alpha$, $\beta$, and
$\gamma$ estimated at the $95\%$ C.L. are still quite large, though
correlated.
 
% This is Figure 3
\begin{figure}
\vskip -1.0truein
\centerline{\epsfxsize 3.5 truein \epsfbox {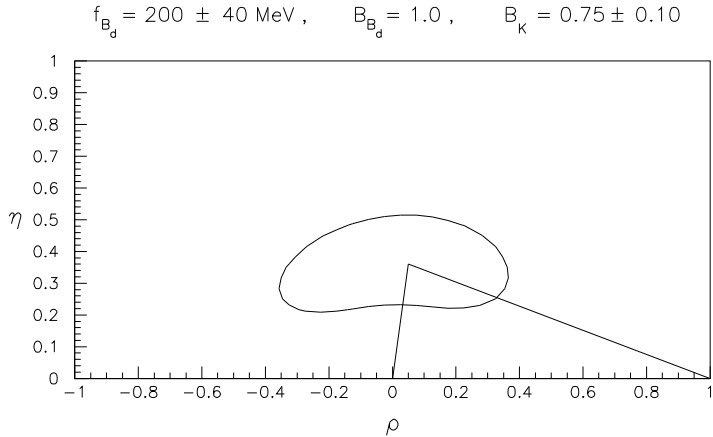}}
\vskip -1.4truein
\caption{Allowed region in $\rho$-$\eta$ space, from a simultaneous fit to
both the experimental and theoretical quantities given in Table
\protect{\ref{tabfit}}. The theoretical errors are treated as Gaussian for
this fit. The solid line represents the region with $\chi^2=\chi_{min}^2+6$
corresponding to the 95\% C.L.\ region. The triangle shows the best fit.}
\label{rhoeta2}
\end{figure}

%%%%%%%%%%%%%%%%%%%%%% SECTION 3 %%%%%%%%%%%%%%%%%%%%%%%%%%%%%%%%

\section{$\delms$ (and $\xs$) and the Unitarity Triangle}

Mixing in the \bsbsbar\ system is quite similar to that in the \bdbdbar\
system. The \bsbsbar\ box diagram is again dominated by $t$-quark exchange,
and the mass difference between the mass eigenstates $\delms$ is given by a
formula analogous to that of Eq.~(\ref{xd}):
\beq
\delms = \frac{G_F^2}{6\pi^2}M_W^2M_{B_s}\left(\fbbs\right)
\hat{\eta}_{B_s} y_t f_2(y_t) \vert V_{ts}^*V_{tb}\vert^2~.
\label{xs}
\eeq
Using the fact that $\vert V_{cb}\vert=\vert V_{ts}\vert$ (Eq.~\ref{CKM}),
it is clear that one of the sides of the unitarity triangle, $\vert
V_{td}/\lambda V_{cb}\vert$, can be obtained from the ratio of $\delmd$ and
$\delms$,
\beq
\frac{\delms}{\delmd} =
 \frac{\hat{\eta}_{B_s}M_{B_s}\left(\fbbs\right)}
{\hat{\eta}_{B_d}M_{B_d}\left(\fbb\right)}
\left\vert \frac{V_{ts}}{V_{td}} \right\vert^2.
\label{xratio}
\eeq
All dependence on the $t$-quark mass drops out, leaving the square of the
ratio of CKM matrix elements, multiplied by a factor which reflects
$SU(3)_{\rm flavour}$ breaking effects. The only real uncertainty in this
factor is the ratio of hadronic matrix elements. Whether or not $\xs$ can
be used to help constrain the unitarity triangle will depend crucially on
the theoretical status of the ratio $\fbbs/\fbb$. In what follows, we will
take $\xi_s \equiv (f_{B_s} \sqrt{\hat{B}_{B_s}}) / (f_{B_d}
\sqrt{\hat{B}_{B_d}}) = (1.15 \pm 0.05)$, consistent with both lattice-QCD
\cite{Wittig96} and QCD sum rules \cite{Narison}. (The SU(3)-breaking
factor in $\delms/\delmd$ is $\xi_s^2$.)

The mass and lifetime of the $B_s$ meson have now been measured at LEP and
Tevatron and their present values are $M_{B_s}=5369.3 \pm 2.0$ MeV and
$\tau(B_s)= 1.52 \pm 0.07 ~ps$ \cite{Richman}. The QCD correction factor
$\hat{\eta}_{B_s}$ is equal to its $B_d$ counterpart, i.e.\
$\hat{\eta}_{B_s} =0.55$. The main uncertainty in $\delms$ (or,
equivalently, $\xs$) is now $\fbbs$. Using the determination of $A$ given
previously, and $\overline{\mt}=165 \pm 9$ GeV, we obtain
\begin{eqnarray}
\delms &=& \left(12.8 \pm 2.1\right)\frac{\fbbs}{(230~\mbox{MeV})^2} 
~(ps)^{-1}~, \nonumber \\
\xs &=& \left(19.5 \pm 3.3\right)\frac{\fbbs}{(230~\mbox{MeV})^2}~.
\end{eqnarray}
The choice $f_{B_s}\sqrt{\hat{B}_{B_s}}= 230$ MeV corresponds to the
central value given by the lattice-QCD estimates, and with this our fits
give $\xs \simeq 20$ as the preferred value in the SM. Allowing the
coefficient to vary by $\pm 2\sigma$, and taking the central value for
$f_{B_s}\sqrt{\hat{B}_{B_s}}$, this gives 
\begin{eqnarray} 
12.9 &\leq & \xs \leq 26.1~, \nonumber\\
8.6 ~(ps)^{-1} &\leq & \delms \leq 17.0 ~(ps)^{-1}~.
 \label{bestxs}
\end{eqnarray}
It is difficult to ascribe a confidence level to this range due to the
dependence on the unknown coupling constant factor. All one can say is that
the standard model predicts large values for $\delms$ (and hence $\xs$).
The present experimental limit $\delms > 9.2
~(ps)^{-1}$ \cite{Gibbons96} is marginally better than the lower bound
on this quantity given above.

An alternative estimate of $\delms$ (or $\xs$) can also be obtained by
using the relation in Eq.~(\ref{xratio}). Two quantities are required.
First, we need the CKM ratio $\vert V_{ts}/V_{td} \vert$. In
Fig.~\ref{vtdts} we show the allowed values (at 95\% C.L.) of the inverse
of this ratio as a function of $\fbd\sqrt{\hat{B}_{B_d}}$, for
$\hat{B}_K=0.75\pm 0.1$. From this one gets
\beq
2.94 \leq \left\vert {V_{ts} \over V_{td}} \right\vert \leq 6.80~.
\eeq
We note in passing that unitarity of the CKM matrix constrains this ratio
to be $3.0 \le |V_{ts}/V_{td}| \le 9.1$ (this can be obtained from
Eq.~\ref{trianglerel}, along with the experimental values of $\lambda$ and
$|V_{ub}/V_{cb}|$.)

The second ingredient is the SU(3)-breaking factor which we take to be
$\xi_s = 1.15 \pm 0.05$, or $1.21 \le \xi_s^2 \le 1.44$. The result of the
CKM fit can therefore be expressed as a $95\%$ C.L. range:
\beq
 11.4 \left(\frac{\xi_s}{1.15}\right)^2
 ~\leq ~\frac{\delms}{\delmd} ~\leq ~
 61.2 \left(\frac{\xi_s}{1.15}\right)^2 ~.
\eeq
Again, it is difficult to assign a true confidence level to $\delms/\delmd$
due to the dependence on $\xi_s$. However, the uncertainty due to the CKM
matrix element ratio has now been reduced to a factor 5.3 due to the
tighter constraints on the unitarity triangle -- our previous fits
\cite{AL95} gave a factor 7.4. The allowed range for the ratio
$\delms/\delmd$ shows that this method is still poorer at present for the
determination of the range for $\delms$, as compared to the absolute value
for $\delms$ discussed above, which in comparison is uncertain by a factor
of 2. Both suffer from additional dependences on
$f_{B_s}\sqrt{\hat{B}_{B_s}}$ or $\xi_s$.

% This is Figure 4
\begin{figure}
\vskip -1.0truein
\centerline{\epsfxsize 3.5 truein \epsfbox {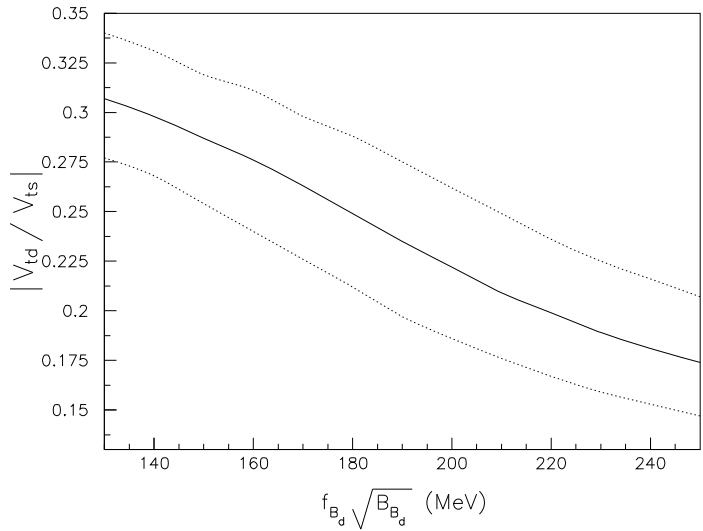}}
\vskip -1.0truein
\caption{Allowed values of the CKM matrix element ratio $\vert
V_{td}/V_{ts} \vert$ as a function of the coupling constant product
$f_{B_d}\protect\sqrt{\hat{B}_{B_d}}$, for $\hat{B}_K=0.75\pm 0.1$. The solid
line corresponds to the best fit values and the dotted curves correspond to
the maximum and minimum allowed values at 95 \% C.L.}
\label{vtdts}
\end{figure}

The present lower bound from LEP $\delms > 9.2~(ps)^{-1}$ (95\% C.L.)
\cite{Gibbons96} and the present world average $\delmd = (0.464
\pm 0.018)~(ps)^{-1}$ can be used to put a bound on the ratio
$\delms/\delmd$, yielding $\delms/\delmd > 19.0$. This  is significantly 
better than the
lower bound on this quantity from the CKM fits, using the central value for
$\xi_s$. The 95\% confidence limit on
$\delms/\delmd$ can be turned into a bound on the CKM parameter space
$(\rho,\eta)$ by choosing a value for the SU(3)-breaking parameter
$\xi_s^2$. We assume three representative values: $\xi_s^2 = 1.21$, $1.32$
and $1.44$, and display the resulting constraints in Fig.~\ref{xslimit}.
From this graph we see that the LEP bound now restricts the allowed
$\rho$-$\eta$ region for all three values of $\xi_s^2$, though this
restriction is weakest for the largest value of $\xi_s^2$ assumed. This
shows that the LEP bound on $\delms$ provides a more stringent lower
bound on the matrix element ratio $|V_{ts}/V_{td}|$ than that obtained from
the CKM fits without this constraint or that following from the unitarity
of the CKM matrix.

% This is Figure 5
\begin{figure}
\vskip -1.0truein
\centerline{\epsfxsize 3.5 truein \epsfbox {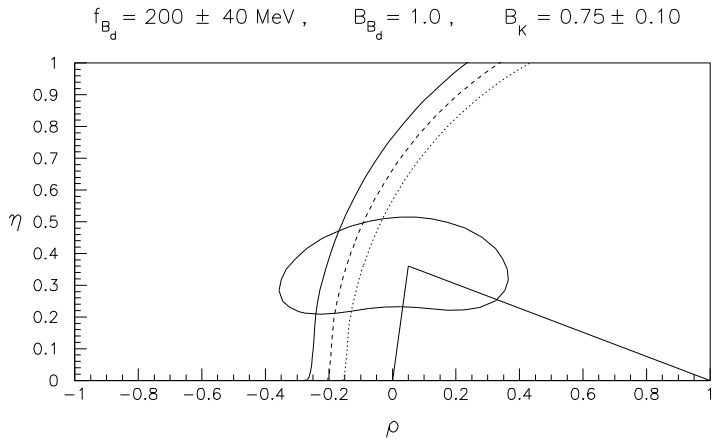}}
\vskip -1.4truein
\caption{Further constraints in $\rho$-$\eta$ space from the LEP bound
 $\delms/\delmd > 19.0$. The bounds are presented for 3 choices of 
the SU(3)-breaking
parameter: $\xi_s^2 = 1.21$ (dotted line), $1.32$ (dashed line) and $1.44$
(solid line). In all cases, the region to the left of the curve is ruled
out.}
\label{xslimit}
\end{figure}

Summarizing the discussion on $\xs$, we note that the lattice-QCD-inspired
estimate $f_{B_s} \sqrt{\hat{B}_{B_s}} \simeq 230$ MeV and the CKM fit
predict that $\xs$ lies between 13 and 26, with a central value around 20.
All of these values scale as $(f_{B_s}\sqrt{\hat{B}_{B_s}}/230
~\mbox{MeV})^2$. The present constraints from the lower bound on $\delms$
on the CKM parameters are now competitive with those from fits to other
data, and this will become even more pronounced with improved data. In
particular, one expects to reach a sensitivity for $\xs \simeq 15$ (or
$\delms \simeq 10~ps^{-1})$ at LEP combining all data and tagging
techniques, and similarly at the SLC, CDF and HERA-B. Of course, an actual
measurement of $\delms$ (equivalently $\xs$) would be very helpful in
further constraining the CKM parameter space. We note that the entire range
for $\xs$ worked out here is accessible at the LHC experiments.

%%%%%%%%%%%%%%%%%%%%%% SECTION 4 %%%%%%%%%%%%%%%%%%%%%%%%%%%%%%%%

\section{CP Violation in the $B$ System}

It is expected that the $B$ system will exhibit large CP-violating effects,
characterized by nonzero values of the angles $\alpha$, $\beta$ and
$\gamma$ in the unitarity triangle (Fig.~\ref{triangle}) \cite{BCPasym}.
The most promising method to measure CP violation is to look for an
asymmetry between $\Gamma(B^0\to f)$ and $\Gamma({\overline{B^0}}\to f)$,
where $f$ is a CP eigenstate. If only one weak amplitude contributes to the
decay, the CKM phases can be extracted cleanly (i.e.\ with no hadronic
uncertainties). Thus, $\sin 2\alpha$, $\sin 2\beta$ and $\sin 2\gamma$ can
in principle be measured in $\bdbarp \to \pi^+ \pi^-$, $\bdbarp\to J/\psi
K_S$ and $\bsbarp\to\rho K_S$, respectively.

Penguin diagrams \cite{penguins} will, in general, introduce some hadronic
uncertainty into an otherwise clean measurement of the CKM phases. In the
case of $\bdbarp\to J/\psi K_S$, the penguins do not cause any problems,
since the weak phase of the penguin is the same as that of the tree
contribution. Thus, the CP asymmetry in this decay still measures $\sin
2\beta$. For $\bdbarp \to \pi^+ \pi^-$, however, although the penguin is
expected to be small with respect to the tree diagram, it will still
introduce a theoretical uncertainty into the extraction of $\alpha$. This
uncertainty can, in principle, be removed by the use of an isospin analysis
\cite{isospin}, which requires the measurement of the rates for
$B^+\to\pi^+\pi^0$, $B^0\to\pi^+\pi^-$ and $B^0\to\pi^0\pi^0$, as well as
their CP-conjugate counterparts. Thus, even in the presence of penguin
diagrams, $\sin 2\alpha$ can in principle be extracted from the decays
$B\to\pi\pi$. Still, this isospin program is ambitious experimentally. If
it cannot be carried out, the error induced on $\sin 2\alpha$ is of order
$|P/T|$, where $P$ ($T$) represents the penguin (tree) diagram. The ratio
$|P/T|$ is difficult to estimate since it is dominated by hadronic physics.
However, one ingredient is the ratio of the CKM elements of the two
contributions: $|V_{tb}^* V_{td} / V_{ub}^* V_{ud} | \simeq
|V_{td}/V_{ub}|$. From our fits, we have determined the allowed values of
$|V_{td}|$ as a function of $|V_{ub}|$. This is shown in Fig.~\ref{vtdvub}
for the ``combined fit". The allowed range for the ratio of these CKM
matrix elements is
\beq
1.4 \leq \left\vert {V_{td}\over V_{ub}} \right\vert \leq 4.6 ~,
\eeq
with a central value of about 3.

% This is Figure 6
\begin{figure}
\vskip -1.0truein
\centerline{\epsfxsize 3.5 truein \epsfbox {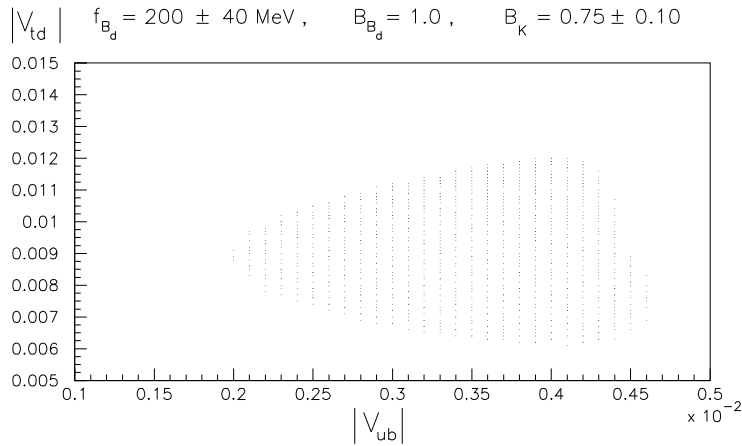}}
\vskip -1.4truein
\caption{Allowed region of the CKM matrix elements $|V_{td}|$ and
$|V_{ub}|$ resulting from the ``combined fit" of the data for the ranges
for $\fbd\protect\sqrt{\hat{B}_{B_d}} $ and $\hat{B}_K$ given in the text.}
\label{vtdvub}
\end{figure}

It is $\bsbarp\to\rho K_S$ which is most affected by penguins. In fact,
the penguin contribution is probably larger in this process than the tree
contribution. This decay is clearly not dominated by one weak (tree)
amplitude, and thus cannot be used as a clean probe of the angle $\gamma$.
Instead, two other methods have been devised, not involving CP-eigenstate
final states. The CP asymmetry in the decay $\bsbarp\to D_s^\pm K^\mp$ can
be used to extract $\sin^2 \gamma$ \cite{ADK}. Similarly, the CP asymmetry
in $B^\pm\to\dcp K^\pm$ also measures $\sin^2 \gamma$ \cite{growyler}.
Here, $\dcp$ is a $D^0$ or $\dbar$ which is identified in a CP-eigenstate
mode (e.g.\ $\pi^+\pi^-$, $K^+K^-$, ...). 

These CP-violating asymmetries can be expressed straightforwardly in terms
of the CKM parameters $\rho$ and $\eta$. The 95\% C.L.\ constraints on
$\rho$ and $\eta$ found previously can be used to predict the ranges of
$\sin 2\alpha$, $\sin 2\beta$ and $\sin^2 \gamma$ allowed in the standard
model. The allowed ranges which correspond to each of the figures in
Fig.~\ref{rhoeta1}, obtained from Fit 1, are found in Table \ref{cpasym1}.
In this table we have assumed that the angle $\beta$ is measured in
$\bdbarp\to J/\Psi K_S$, and have therefore included the extra minus sign
due to the CP of the final state.

\begin{table}
\hfil
\vbox{\offinterlineskip
\halign{&\vrule#&
 \strut\quad#\hfil\quad\cr
\noalign{\hrule}
height2pt&\omit&&\omit&&\omit&&\omit&\cr
& $\fbd\sqrt{\hat{B}_{B_d}}$ (MeV) && $\sin 2\alpha$ &&
$\sin 2\beta$ && $\sin^2 \gamma$ & \cr
height2pt&\omit&&\omit&&\omit&&\omit&\cr
\noalign{\hrule}
height2pt&\omit&&\omit&&\omit&&\omit&\cr
& $160$ && 0.86 -- 1.0 && 0.38 -- 0.58 && 0.47 -- 0.92 & \cr
& $180$ && $-$0.05 -- 1.0 && 0.47 -- 0.71 && 0.77 -- 1.0 & \cr
& $200$ && $-$0.69 -- 0.90 && 0.53 -- 0.81 && 0.61 -- 1.0 & \cr
& $220$ && $-$0.85 -- 0.61 && 0.60 -- 0.89 && 0.44 -- 0.98 & \cr
& $240$ && $-$0.90 -- 0.32 && 0.67 -- 0.94 && 0.34 -- 0.86 & \cr
height2pt&\omit&&\omit&&\omit&&\omit&\cr
\noalign{\hrule}}}
\caption{The allowed ranges for the CP asymmetries $\sin 2\alpha$, $\sin
2\beta$ and $\sin^2 \gamma$, corresponding to the constraints on $\rho$ and
$\eta$ shown in Fig.~\protect\ref{rhoeta1}. Values of the coupling constant
$\fbd\protect\sqrt{\hat{B}_{B_d}}$ are stated. We fix $\hat{B}_K=0.75$. The
range for $\sin 2\beta$ includes an additional minus sign due to the CP of
the final state $J/\Psi K_S$.}
\label{cpasym1}
\end{table}

% This is Figure 7
\begin{figure}
\vskip -2.4truein
\centerline{\epsfxsize 7.0 truein \epsfbox {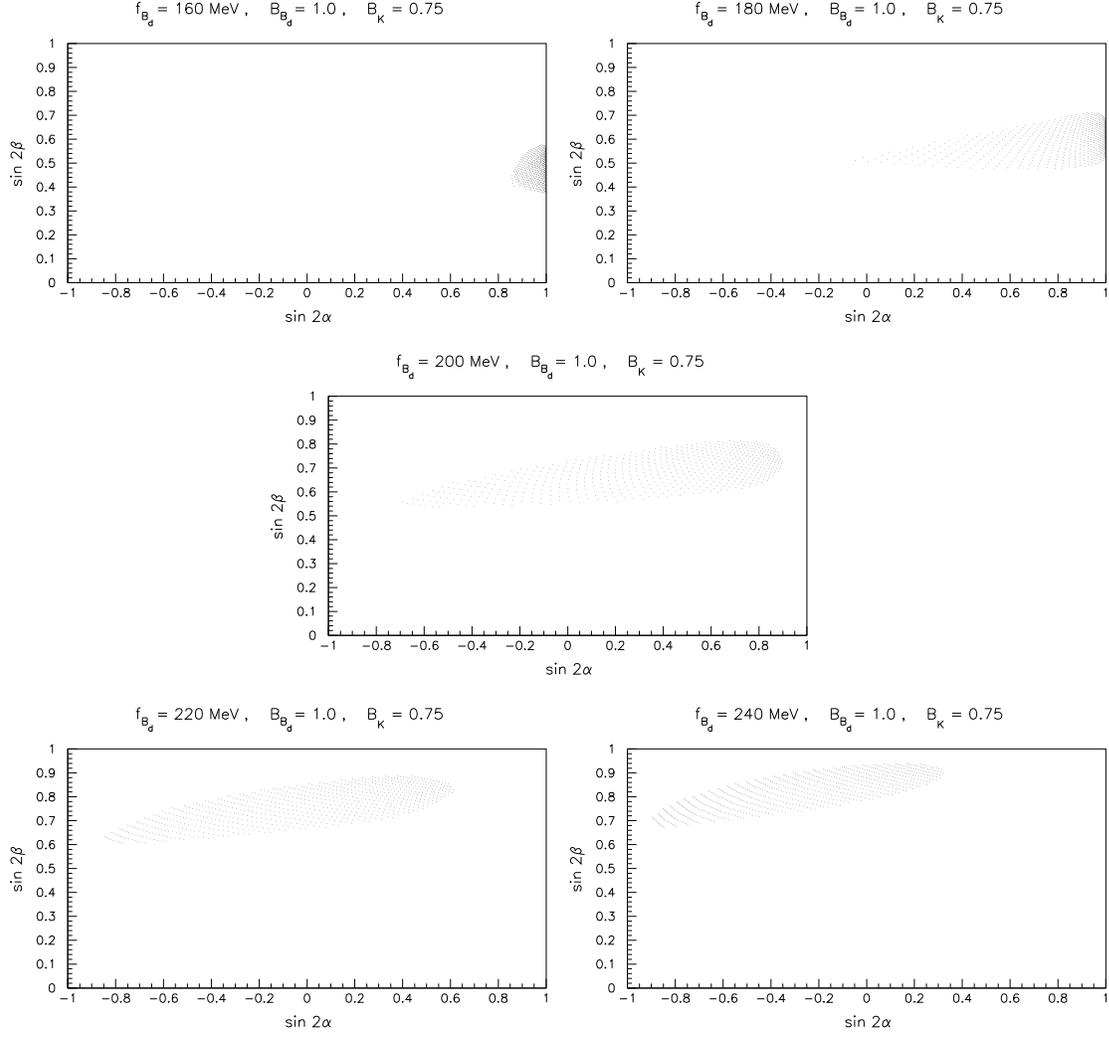}}
\vskip -1.0truein
\caption{Allowed region of the CP asymmetries $\sin 2\alpha$ and $\sin
2\beta$ resulting from the ``experimental fit" of the data for different
values of the coupling constant $\fbd\protect\sqrt{\hat{B}_{B_d}}$
indicated on the figures a) -- e). We fix $\hat{B}_K=0.75$.}
\label{alphabeta1}
\end{figure}

Since the CP asymmetries all depend on $\rho$ and $\eta$, the ranges for
$\sin 2\alpha$, $\sin 2\beta$ and $\sin^2 \gamma$ shown in Table
\ref{cpasym1} are correlated. That is, not all values in the ranges are
allowed simultaneously. We illustrate this in Fig.~\ref{alphabeta1},
corresponding to the ``experimental fit" (Fit 1), by showing the region in
$\sin 2\alpha$-$\sin 2\beta$ space allowed by the data, for various values
of $\fbd\sqrt{\hat{B}_{B_d}}$. Given a value for
$\fbd\sqrt{\hat{B}_{B_d}}$, the CP asymmetries are fairly constrained.
However, since there is still considerable uncertainty in the values of the
coupling constants, a more reliable profile of the CP asymmetries at
present is given by our ``combined fit" (Fit 2), where we convolute the
present theoretical and experimental values in their allowed ranges. The
resulting correlation is shown in Fig.~\ref{alphabeta2}. From this figure
one sees that the smallest value of $\sin 2\beta$ occurs in a small region
of parameter space around $\sin 2\alpha\simeq 0.8$-0.9. Excluding this
small tail, one expects the CP-asymmetry in $\bdbarp\to J/\Psi K_S$ to be
at least 20\% (i.e., $\sin 2 \beta > 0.4)$. Finally, in the SM the
relation $\alpha+\beta+\gamma=\pi$ is satisfied. However, note that the
allowed range for $\beta$ is rather small (Table \ref{cpasym1}). Thus,
there should be a strong correlation between $\alpha$ and $\gamma$
\cite{DGR}. This is indeed the case, as is shown in Fig.~\ref{alphagam}.

% This is Figure 8
\begin{figure}
\vskip -1.0truein
\centerline{\epsfxsize 3.5 truein \epsfbox {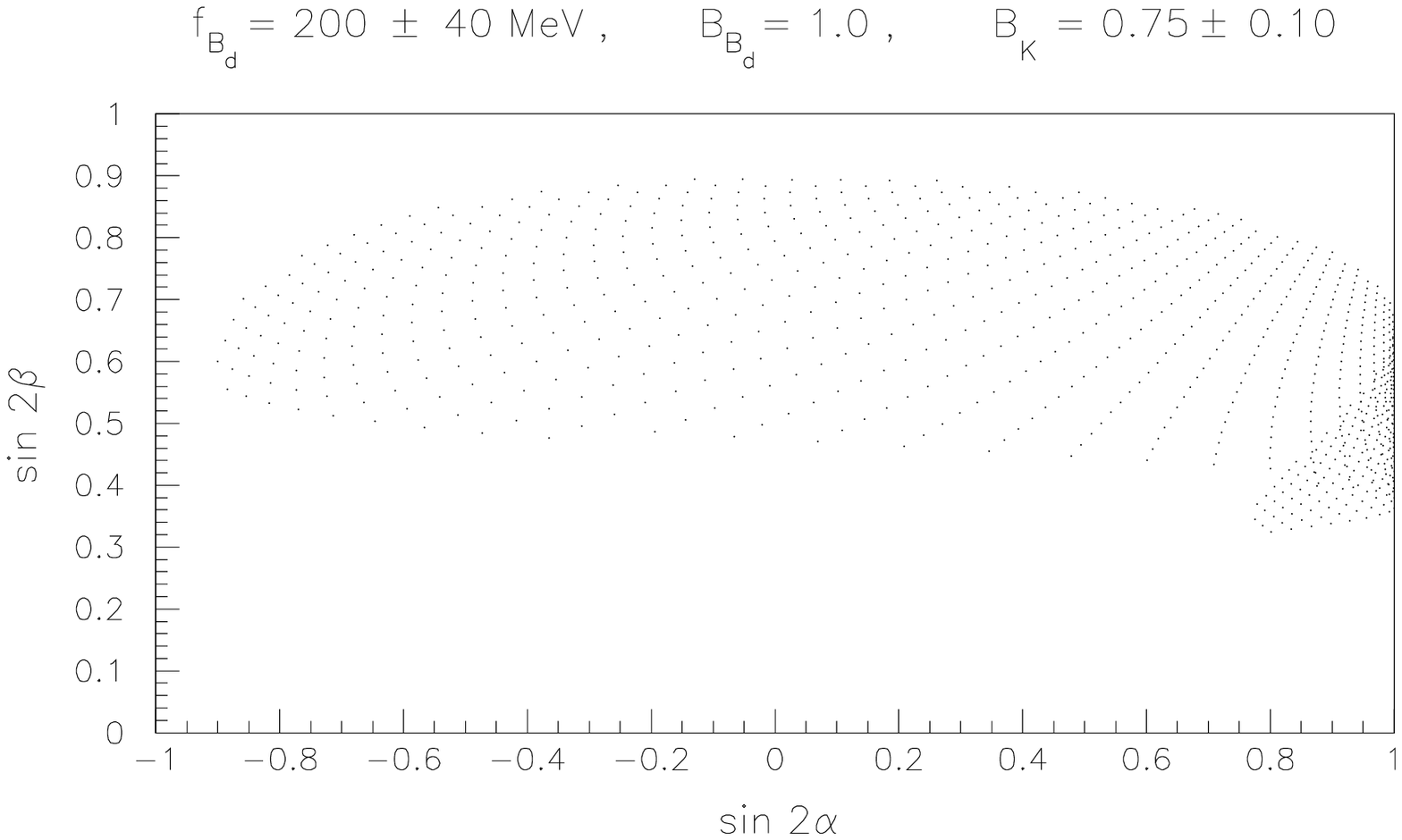}}
\vskip -1.4truein
\caption{Allowed region of the CP-violating quantities $\sin 2\alpha$ and 
$\sin 2\beta$ resulting from the ``combined fit" of the data for the ranges
for $\fbd\protect\sqrt{\hat{B}_{B_d}} $ and $\hat{B}_K$ given in the text.}
\label{alphabeta2}
\end{figure}

% This is Figure 9
\begin{figure}
\vskip -1.0truein
\centerline{\epsfxsize 3.5 truein \epsfbox {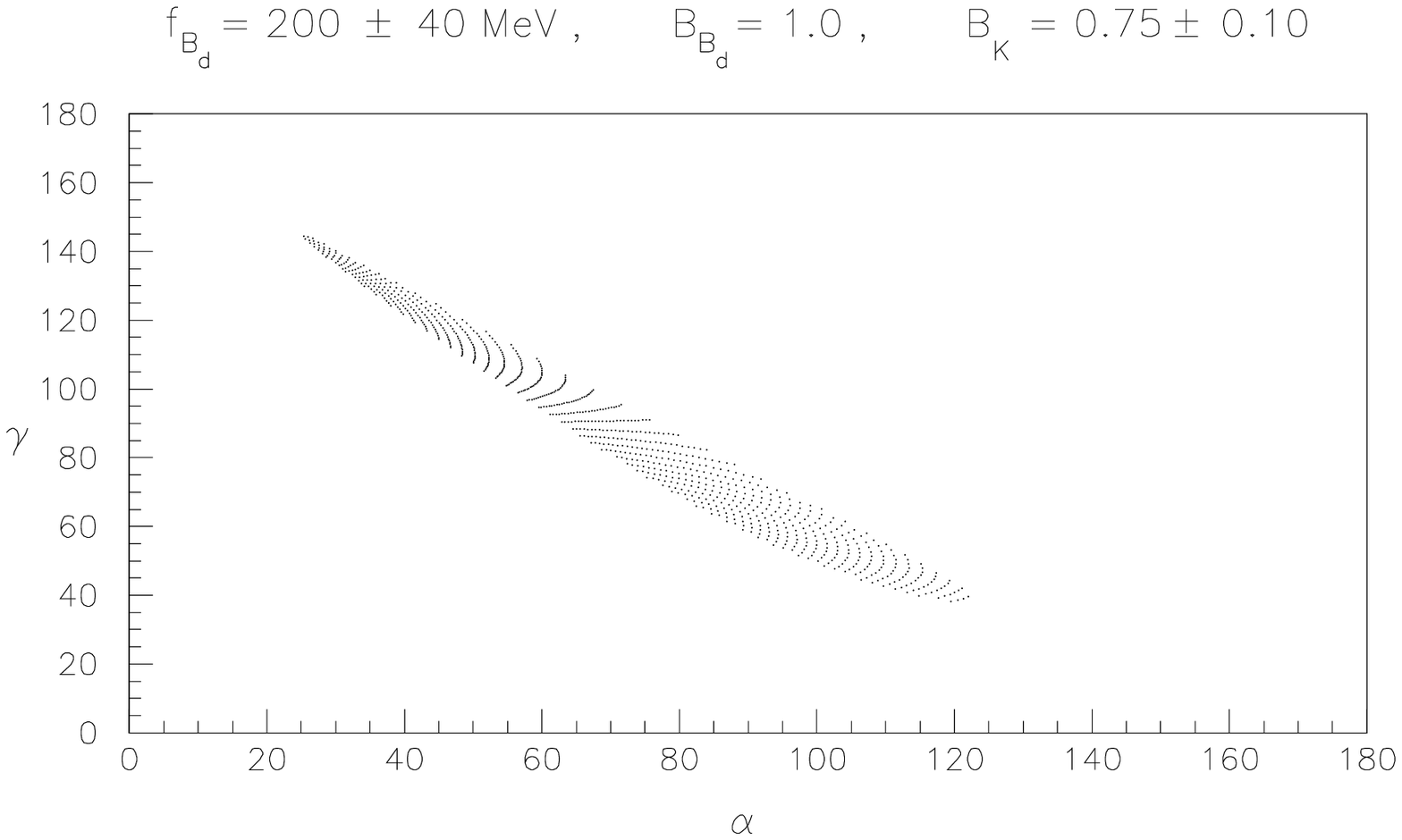}}
\vskip -1.4truein
\caption{Allowed values (in degrees) of the angles $\alpha$ and $\gamma$ 
resulting from the ``combined fit" of the data for the ranges for 
$\fbd\protect\sqrt{\hat{B}_{B_d}} $ and $\hat{B}_K$ given in the text.} 
\label{alphagam}
\end{figure}

%%%%%%%%%%%%%%%%%%%%%% SECTION 5 %%%%%%%%%%%%%%%%%%%%%%%%%%%%%%%%

\section{Summary and Outlook}

We summarize our results:

\smallskip

(i) We have presented an update of the CKM unitarity triangle using the
theoretical and experimental improvements in the following quantities:
$\abseps$, $\absvcb$, $\vert V_{ub}/V_{cb} \vert$, $\delmd$, $\tau(B_d)$,
${\overline{\mt}}$. The fits can be used to exclude extreme values of the
pseudoscalar coupling constants, with the range $130~\mbox{MeV} < f_{B_d}
\sqrt{\hat{B}_{B_d}} < 250~\mbox{MeV}$ still allowed for $\hat{B}_K=0.75$,
although the fits at the two boundary values of $ f_{B_d}
\sqrt{\hat{B}_{B_d}}$ are poor.

\smallskip

(ii) The newest experimental and theoretical numbers restrict the allowed
CKM unitarity triangle in the $(\rho,\eta)$-space considerably more than
before. However, the present uncertainties are still large -- despite the
new, more accurate experimental data, our knowledge of the unitarity
triangle is still deficient. This underscores the importance of measuring
CP-violating rate asymmetries in the $B$ system. Such asymmetries are
largely independent of theoretical hadronic uncertainties, so that their
measurement will allow us to accurately pin down the parameters of the CKM
matrix. Furthermore, unless our knowledge of the pseudoscalar coupling
constants improves considerably, better measurements of such quantities as
$\xd$ will not help much in constraining the unitarity triangle. On this
point, help may come from the experimental front. It may be possible to
measure the parameter $\fbd$, using isospin symmetry, via the
charged-current decay $\bu\to\tau^\pm \nu_\tau$. Along the same lines, the
prospects for measuring $(\fbd,\fbs)$ in the FCNC leptonic and photonic
decays of $\bd $ and $\bs$ hadrons, $(\bd,\bs)\to\mu^+\mu^-, (\bd,\bs) \to
\gamma\gamma$ in future $B$ physics facilities are not entirely dismal
\cite{ALI96}.

\smallskip

(iii) We have determined bounds on the ratio $\vert V_{td}/V_{ts} \vert$
from our fits. For $130~\mbox{MeV} \leq f_{B_d} \sqrt{\hat{B}_{B_d}} \leq
250~\mbox{MeV}$, i.e.\ in the entire allowed domain, at 95 \% C.L. we find
\beq
0.15 \leq \left\vert {V_{td} \over V_{ts}} \right\vert \leq 0.34~.
\eeq
 These bounds are now better than those obtained
from unitarity, which gives $0.11 \leq \vert V_{td}/V_{ts} \vert \leq
0.33$.
Furthermore, the upper bound from our analysis is more restrictive than the 
current
experimental upper limit following from the CKM-suppressed radiative
penguin decays $BR(B \to \omega + \gamma )$ and $BR(B \to \rho + \gamma )$,
which at present yield at 90\% C.L. \cite{cleotdul}
\beq
\left\vert {V_{td} \over V_{ts}} \right\vert \leq 0.45 - 0.56~,
\eeq
depending on the model used for the SU(3)-breaking in the relevant form
factors \cite{SU3ff}. Long-distance effects in the decay $B^\pm \to
\rho^\pm + \gamma$ may introduce theoretical uncertainties comparable to
those in the SU(3)-breaking part but the corresponding effects in the
decays $B^0 \to (\rho^0,\omega) +\gamma$ are expected to be very small
\cite{ab95}.

\smallskip

(iv) Using the measured value of $\mt$, we find
\begin{equation}
\xs = \left(19.5 \pm 3.3\right)\frac{\fbbs}{(230 ~\mbox{MeV})^2}~.
\end{equation}
Taking $f_{B_s}\sqrt{\hat{B}_{B_s}}= 230$ (the central value of lattice-QCD
estimates), and allowing the coefficient to vary by $\pm 2\sigma$, this
gives
\begin{equation}
12.9 \leq \xs \leq 26.1~.
\end{equation}
No reliable confidence level can be assigned to this range -- all that one
can conclude is that the SM predicts large values for $\xs$, which lie mostly
above the LEP 95\% C.L. lower limit of $\delms > 9.2 ~(ps)^{-1}$, which
on using $\tau(B_s)=1.52~ps$ gives $\xs > 14.0$.

\smallskip

(v) The ranges for the CP-violating rate asymmetries parametrized by $\sin
2\alpha$, $\sin 2\beta$ and and $\sin^2 \gamma$ are determined at 95\% C.L.
to be
\begin{eqnarray}
&~& -0.90 \leq \sin 2\alpha \le 1.0~, \nonumber \\
&~& 0.32 \leq \sin 2\beta \le 0.94~, \\
&~& 0.34 \leq \sin^2 \gamma \le 1.0~. \nonumber
\end{eqnarray}
Penguin amplitudes may play a significant role in some methods of
extracting the CKM phases. Their magnitude, relative to the tree
contribution, is therefore of some importance. One factor in determining
this relative size is the ratio of CKM matrix elements $\vert V_{td}/V_{ub}
\vert$. We find
\beq
1.4 \leq \left\vert {V_{td}\over V_{ub}} \right\vert \leq 4.6 ~.
\eeq

\bigskip
\noindent
{\bf Acknowledgements}:
\bigskip

We thank Roger Forty, Lawrence Gibbons, Guido Martinelli, Hans-Guenther 
Moser, Stephan Narison, Sheldon Stone, Ed Thorndike and Christian Zeitnitz
for very helpful discussions. Some experimental results used in these CKM
fits have been kindly communicated to us by Lawrence Gibbons. A.A. thanks
Stephan Narison for the hospitality in Montpellier during the conference
QCD '96.

%\vskip0.5truein

%%%%%%%%%%%%%%%%%%%%% REFERENCES %%%%%%%%%%%%%%%%%%%%%%%%%%%%%%%%


\begin{thebibliography}{99}

\bibitem{CKM} N. Cabibbo, Phys.\ Rev.\ Lett.\ {\bf 10} (1963) 531; M.
Kobayashi and K. Maskawa, Prog.\ Theor.\ Phys.\ {\bf 49} (1973) 652.

\bibitem{AL95} A. Ali and D. London, preprint DESY 95-148,
UdeM-GPP-TH-95-32 [hep-ph/9508272], to appear in the {\it  Proc.\ of the
6th Int.\ Symp.\ on Heavy Flavour Physics}, Pisa, June 6-9, 1995.

\bibitem{Wolfenstein} L. Wolfenstein, Phys.\ Rev.\ Lett.\ {\bf 51} (1983)
1945.

\bibitem{AL94} A. Ali and D. London, \zpc{65}{95}{431}.

\bibitem{PDG96} R.M. Barnett et al.\ (Particle Data Group), Phys.\ Rev.\
{\bf D54} (1996) 1.

\bibitem{DPG96} Th.\ M\"uller, Plenar Votrag auf der Fr\"uhjahrstagung 1996
der deutschen Physikalischen Gesellschaft - Sektion Teilchenphysik,
Hamburg, FRG, M\"arz 18-21, 1996.
 
\bibitem{mtmsbar} N. Gray, D.J. Broadhurst, W. Grafe, and K. Schilcher, Z.
Phys.\ {\bf C48} (1990) 673.

\bibitem{neubert95} M. Neubert, Preprint CERN-TH/95-107 [hep-ph/9505238].

\bibitem{shifman95} M. Shifman, Preprint TPI-MINN-95/15-T [hep-ph/9505289].

\bibitem{Tomasz95} T. Skwarnicki, preprint [hep-ph/9512395], to appear in
the {\it Proc.\ of the 17th Int.\ Symp.\ on Lepton Photon Interactions},
Beijing, P.R. China, August 1995.

\bibitem{Wisgur} N. Isgur and M.B. Wise, \plb{232}{89}{113}; 237 (1990)
527.

\bibitem{Gibbons96} L. Gibbons (CLEO Collaboration), Invited talk at the
International Conference on High Energy Physics, Warsaw, ICHEP96 (1996).

\bibitem{Cz96} A. Czarnecki, Phys.\ Rev.\ Lett.\ {\bf 76} (1996) 4124.

\bibitem{Luke} M.E. Luke, \plb{252}{90}{447}.

\bibitem{GKLW96} M. Gremm, A. Kapustin, Z. Ligeti, and M.B. Wise, preprint
CALT-68-2043 [hep-ph/9603314].

\bibitem{Bartelt93} J. Bartelt et al.\ (CLEO Collaboration), Phys.\ Rev.\
Lett.\ {\bf 64} (1990) 16.

\bibitem{Burasetal} A.J. Buras, W. Slominski, and H. Steger, Nucl.\ Phys.\
{\bf B238} (1984) 529; {\it ibid.} {\bf B245} (1984) 369.

\bibitem{HN94} S. Herrlich and U. Nierste, Nucl.\ Phys.\ {\bf B419} (1994)
292.

\bibitem{etaB} A.J. Buras, M. Jamin and P.H. Weisz, Nucl.\ Phys.\ {\bf
B347} (1990) 491.

\bibitem{HN95} S. Herrlich and U. Nierste, Phys.\ Rev.\ {\bf D52} (1995)
6505.

\bibitem{Soni95} A. Soni, preprint [hep-lat/9510036] (1995).

\bibitem{BP95} J. Bijnens and J. Prades, Nucl.\ Phys.\ {\bf B444} (1995)
523.

\bibitem{Sharpe94} S. Sharpe, Nucl.\ Phys.\ B (Proc.\ Suppl.) {\bf 34}
(1994) 403.

\bibitem{Crisafulli95} M. Crisafulli et al.\ (APE Collaboration), Phys.\
Lett.\ {\bf B369} (1996) 325.

\bibitem{JLQCD} S. Aoki et al.\ (JLQCD Collaboration), preprint UTHEP-322
(1995) [hep-lat/9510012]. The numbers cited for $B_K$ from the JLQCD
collaboration as well as from the work of Soni and Bernard are quoted by
Soni in his review \protect\cite{Soni95}.

\bibitem{Zeitnitz} C. Zeitnitz, invited talk at the 4th International
Workshop on $B$-Physics at Hadron Machines (BEAUTY 96), Rome, 17-21 June,
1996, and private communication.

\bibitem{Wittig96} H. Wittig, Invited talk presented at the III
German-Russian Workshop on Heavy Quark Physics, Dubna, Russia, 20-22 May,
1996, [hep-ph/9606371].

\bibitem{UKQCDBB} A.K. Ewing et al.\ (UKQCD Collaboration), preprint,
[hep-lat-9508030] (1995).
 
\bibitem{Petronzio96}  R. Petronzio (ALPHA Collaboration), private
communication.

\bibitem{narison95} S. Narison, Phys.\ Lett.\ {\bf B351} (1995) 369.

\bibitem{Saulanwu95} Sau Lan Wu, preprint WISC-EX-96-343 [hep-ex/9602003],
to appear in the {\it Proc.\ of the 17th Int.\ Symp.\ on Lepton and Photon
Interactions}, Beijing, P.R. China, August 1995.

\bibitem{ALEPHxs96} ``Combined limit on the $B_s^0$ oscillation frequency",
contributed paper by the ALEPH collaboration to the International
Conference on High Energy Physics, Warsaw, ICHEP96 PA08-020 (1996).

\bibitem{BCPasym} For reviews, see, for example, Y. Nir and H.R. Quinn, in
{\it $B$ Decays}, edited by S. Stone (World Scientific, Singapore, 1992)
362; I. Dunietz, {\it ibid} 393.

\bibitem{AKL94} R. Aleksan, B. Kayser, and D. London, Phys.\ Rev.\ Lett.\
{\bf B73} (1994) 18.

\bibitem{Narison} S. Narison, Phys.\ Lett.\ {\bf B322} (1994) 247; S.
Narison and A. Pivovarov, {\it ibid} {\bf B327} (1994) 341.

\bibitem{Richman} J. Richman, plenary talk at the International Conference
on High Energy Physics, Warsaw, ICHEP96 (1996).

\bibitem{penguins} D. London and R. Peccei, Phys.\ Lett.\ {\bf B223} (1989)
257; B. Grinstein, Phys.\ Lett.\ {\bf B229} (1989) 280; M. Gronau, Phys.\
Rev.\ Lett.\ {\bf 63} (1989) 1451,  Phys.\ Lett.\ {\bf B300} (1993) 163.

\bibitem{isospin} M. Gronau and D. London, Phys.\ Rev.\ Lett.\ {\bf 65}
(1990) 3381.

\bibitem{ADK} R. Aleksan, I. Dunietz, and B. Kayser, Z. Phys.\ {\bf C54}
(1992) 653.

\bibitem{growyler} M. Gronau and D. Wyler, Phys.\ Lett.\ {\bf B265} (1991)
172. See also M. Gronau and D. London, Phys.\ Lett.\ {\bf B253} (1991) 483; 
I. Dunietz, Phys.\ Lett.\ {\bf B270} (1991) 75.

\bibitem{DGR} A.S. Dighe, M. Gronau and J.L. Rosner, CERN-TH/96-68,
hep-ph/9604233, to be published in {\it Phys.~Rev.} {\bf D}.

\bibitem{ALI96} A. Ali, preprint DESY 96-106 [hep-ph/9606324]; to appear in
the Proceedings of the XX International Nathiagali Conference on Physics
and Contemporary Needs, Bhurban, Pakistan, June 24-July 13, 1995 (Nova
Science Publishers, New York, 1996).

\bibitem{cleotdul} R. Ammar et al.\ (CLEO Collaboration), CLEO CONF 96-05,
ICHEP96 PA05-093 (1996).

\bibitem{SU3ff} A. Ali, V.M. Braun and H. Simma, Z. Phys.\ {\bf C63} (1994)
437; J.M. Soares, Phys.\ Rev.\ {\bf D49} (1994) 283; S. Narison, Phys.\
Lett.\ {\bf B327} (1994) 354.

\bibitem{ab95} A. Ali and V.M. Braun, Phys.\ Lett.\ {\bf B359} (1995) 223;
A. Khodzhamirian, G. Stoll, and D. Wyler, Phys.\ Lett.\ {\bf B358} (1995)
129.

\end{thebibliography}
\end{document}